# On a Possible Melting Curve of $C_{60}$ Fullerite

V.I.Zubov[*,1], C.G.Rodrigues[2], and I.V.Zubov[1]

[1] Instituto de Física, Universidade Federal de Goiás, C.P. 131, 74001 Goiânia - GO, Brazil[1] and Department of Theoretical Physics, People's Friendship University, Moscow, Russia

[2] Núcleo de Pesquisa em Física da Universidade Católica de Goiás, Goiânia -GO, Brazil

We study thermodynamic properties of the high-temperature modification of fullerites on the basis of the Girifalco intermolecular potential. In the present work, using the Lindemann's melting criterion we estimate a possible melting curve $T_m(P)$ of $C_{60}$ fullerite. To take into account the lattice anharmonicity that has strong effect at $T > 700$ K, we use the correlative method of unsymmetrized self-consistent field. To check this approach, first we have applied it to solid Ar In the range between its triple point $T_t = 83.807$ K and 260 K and obtained the mean square relative deviation from experimental data about 0.7 %. The melting curve for $C_{60}$ fullerite has been calculated from the melting point at normal pressure estimated at 1500 K up to 15 kbar which corresponds to $T_m = 4000$ K, i.e. to the temperature estimated be Kim and Tománek [Phys. Rev. Lett. **72**, 2418 (1994)] as that of the decomposition of the $C_{60}$ molecule itself. The temperature dependence of the melting pressure is approximated very well by the Simon equation $(P_m(T)/\text{bar} - 1)/a = (T/T_0)^c$ with $T_0 = 1500$ K, $a = 6643.8$, $c = 1.209$. The temperature dependence of the molar volume along the melting curve is discribed by $V_s(T) = V_s(T_0) - 29.20 \ln(T/T_0)$.

Мы изучаем термодинамические свойства высокотемпературной модификации фуллеритов на основе межмолекулярного потенциала Жирифалко. В настоящей работе, используя критерий плавления Линдеманна, оценивается возможная линия плавления $T_m(P)$ фуллерита $C_{60}$. Для учета ангармонизма колебаний решетки, который является сильным при $T > 700$ K, используется корреляционный метод несимметризованного самосогласованного поля. Для проверки используемого подхода он сначала был применен к твердому Ar. В интервале температур от тройной точки $T_t = 83.807$ K до 260 К получено среднеквадратичное отклонение от экспериментальных данных 0.7 %. Линия плавления фуллерита $C_{60}$ рассчитана от точки плавления при нормальном давлении, оцененной в 1500 К, до 15 кбар, что соответствует 4000 К, оцененной Кимом и Томанеком [Phys. Rev. Lett. **72**, 2418 (1994)] температуры распада самой молекулы. Температурная зависимость давления на линии плавления очень хорошо аппроксимируется уравнением Симона $\left(P_m(T)/\text{bar} - 1\right)/a = \left(T/T_0\right)^c$ с $T_0$ = 1500 K, $a$ = 6643.8, $c$ = 1.209. Температурная зависимость молярного объема вдоль этой линии описывается формулой $V_s(T) = V_s(T_0) - 29.20\ln(T/T_0)$

---

[*] Corresponding author: E-mail v_zubov@mail.ru; zubov@fis.ugf.br

## 1. Introduction

Since the discovery of the fullerenes [1] and especially after the elaboration of effective methods for their preparation, separation and purification [2], that gave rise to their production in sufficient quantities for the growth of macroscopic crystals named the fullerites, they have been attracting a great attention of scientists [3 – 6]. Among other things, the thermodynamic properties of fullerites have been investigated. Most of them are due to the lattice vibrations, whereas the dominant contribution to their specific heats comes from the intramolecular degrees of freedom. Immediately after the preparation of fullerite crystals, a start has been made on the explorations of the phase transitions in these materials. The fullerene of the greatest abundance, $C_{60}$ and the next, $C_{70}$ have been studied most intensively. The phase transitions between the low-temperature, orientationally ordered, and high-temperature, disordered, modifications have investigated both experimentally, *e.g.* [7 – 12], and theoretically [13]. The measurements of the saturated vapor pressures and the enthalpies of sublimation of these fullerites and also those of $C_{76}$ and $C_{84}$ have been fulfilled, and Markov *et al*. [14]. have published summary data, see also refs. there. In [15 – 17] one can find theoretical results.

The liquid phase of fullerites has never been observed. Nevertheless, discussion about its possible existence of has persisted for years [18 – 27], based on a cluster-expansion-type method [18], an integral-equation approach [19], a Monte Carlo technique [20, 25, 26], a density-functional theory [21, 23], molecular dynamics simulations [19, 24], a modified hypernetted-chain method [22] and also on the scaling of Lennard-Jones values [15, 19]. Hagen *et al*. [20] have reasoned that the liquid phase of $C_{60}$ has no the region of absolute stability and hence cannot exist. However other authors have drawn the possibility of its liquid phase, although in a narrow phase diagram range. The estimations of its melting temperature (triple point) vary from 1400 to 1800 K. In our opinion, upper values are closer to the spinodal point of the solid phase rather than to the melting temperature. Hasewaga and Ohno [25, 26] have also evaluated the liquid-vapor binodal curve including the critical point. They and Ferreira et al. [27] have also studied the solid-liquid coexisting but in a very narrow temperature interval. Note that Stetzer *et al.* [28] have reported that $C_{60}$ crystals heated at 1260 K for more than 10 min decomposed into amorphous carbon. However, this result has not been reproduced in other institutions, whereas the molecular dynamics estimation for the decomposition temperature of a single $C_{60}$ molecule comprises yields about 4000 K [29]. Because of this, the investigations of the possibility of the liquid phase for fullerities are hitherto being continued [24, 27, 29]. Recently Abramo *et al.* [30] have computed the phase diagrams for higher fullerites $C_{70}$, $C_{76}$, $C_{84}$ and $C_{96}$.

The present work is concerned with evaluation for the pressure dependence of rhe possible melting temperature of $C_{60}$ fullerite up to 4000 K.

## 2. Calculation Procedure

Owing to the lack of unified rigorous microscopic theory for crystals and liquids, semi-empirical criteria for melting are of frequent use which are stated as a constancy of one or other characteristics of phase on the melting curve (a peculiar kind of "integrals of movement along the melting curve"). The first to be found was the Lindemann's criterion, see *e.g.* [31]. It implies that on the melting curve

$$\delta = \sqrt{\overline{\vec{q}^2}}/a = const, \qquad (1)$$

where $\overline{\vec{q}^2} = 3\overline{q_\alpha^2}$ is the variance of the molecular positions near their lattice points and *a* the nearest-neighbor distance. Although more recently other criteria have been established (the Ross' criterion [32], the entropy [33] and energy [34] rules), interest has been shown in the Lindemann's criterion until to present time [35].

Here we applied it to the possible melting curve of $C_{60}$ fullerite. At $T > 700$ K, the lattice anharmonicity has a strong effect on its properties. To take it into account we use the correlative method of unsymmetrized self-consistent field (CUSF) [36]. Including anharmonic terms up to the fourth order in the zeroth approximation, the equation of state of the crystal at the temperature $\Theta = kT$ under the pressure *P* is of the form

$$P = -\frac{a}{3v}\left[\frac{1}{2}\frac{dK_0}{da} + \frac{\beta\Theta}{2K_2}\frac{dK_2}{da} + \frac{(3-\beta)\Theta}{4K_4}\frac{dK_4}{da}\right] + P^2 + P^H. \qquad (2)$$

Here $v(a)$ is the volume of the unit cell,

$$K_{2l} = \frac{1}{2l+1}\sum_{k\geq 1} Z_k \nabla^{2l}\Phi(R_k), l=0,1,2, \qquad (3)$$

$\Phi(r)$ is the intermolecular potential, $Z_k$ are the coordinational numbers, $R_k$ are the radii of the coordinational spheres, $\beta\left(K_2\sqrt{3/\Theta K_4}\right)$ is the solution of the transcendental equation

$$\beta = 3X \frac{D_{-2.5}(X+5\beta/6X)}{D_{-1.5}(X+5\beta/6X)}, \tag{4}$$

in which $D_\nu$ are the parabolic cylinder functions, $P^2$ and $P^H$ are corrections of the perturbation theory that turn more accurate the effect of the anharmonicity.

For the intermolecular forces we use the Girifalco potential [37]

$$\Phi_G(r) = -\alpha\left(\frac{1}{s(s-1)^3} + \frac{1}{s(s+1)^3} - \frac{2}{s^4}\right) + \beta\left(\frac{1}{s(s-1)^9} + \frac{1}{s(s+1)^9} - \frac{2}{s^{10}}\right) \tag{5}$$

where $s = r/2a$, $a = 3.55 \times 10^{-8}\,cm$, $\alpha = 7.494 \times 10^{-14}\,erg$ and $\beta = 1.3595 \times 10^{-16}\,erg$. It has the minimum point $r_0 = 10.0558\,E$ and the depth of the potential well is $\varepsilon/k = 3218.4\,K$.

The variance of the molecular positions in strongly anharmonic crystals is expressed in terms of $\Theta$, the derivatives of the intermolecular potential and $\beta(X)$ [37]. The value of the Lindemann's parameter (1) computed at a single melting point $P$, $T_m(P)$, $a(P, T_m)$ that is considered to be known, can de utilized for calculations of the melting curve.

### 3. Results and discuccion

To check the accuracy of the Lindemann's criterion (1) for strongly anharmonic crystals, previously we have applied it to solid Ar since its melting curve is well known from experiments [33]. In the range between its triple point $T_t = 83.807\,K$ and 260 K, the mean square relative deviation from experimental data makes up about 0.7 %.

Using our estimation for the melting temperature of $C_{60}$ fullerite at normal pressure [36], $T_0 = 1500\,K$, we calculated the Lindemann parameter at this point: $\delta \approx 0.041$. Then, we solved the equation of state (2) at various fixed pressures up to temperature $T_m(P)$ at which $\sqrt{\overline{q^2}}/a = 0.041$ [38] and calculated the molar volume at this melting point $V_S = V(P, T_m)$. We have restricted ourselves to a temperature about 4000 K (and a pressure about 15 kbar) since at such temperature the C60 molecule is decomposed [29].

Note that in the quasi-harmonic approximation, the mean-square molecular displacements are somewhat different than taking into account the strong anharmonicity. But such an approximation fails at high temperatures. It will suffice to mention [15] that for the temperature of the loss of thermodynamic stability of the two-phase system C60 fullerite–vapor, $T_S$, it gives 1054

K, which is much below the estimations for its triple point. Besides, this fullerite was observed at higher temperatures. At the same time Eq. (2) yields $T_S \approx 1916$ K.

The results of our calculations are shown in Fig. 1. The temperature dependence of the melting pressure is described very well by the Simon equation

$$\frac{(P_m(T)/\text{bar})-1}{a} = \left(\frac{T}{T_0}\right)^c. \qquad (6)$$

Originally, this equation has been proposed for the melting curve of Ar (see e.g. [33]). For $C_{60}$ fullerite we find $T_0 = 1500$ K, $b = 6643.8$, $c = 1.209$. The temperature dependence of the molar volume along the melting curve is approximated by the formula

$$V_s(T) = V_s(T_0) - 29.20 \ln(T/T_0) \qquad (7)$$

It has been demonstrated with Ar [39] that the logarithmic relationship between the molar volume of the solid phase and the temperature along its melting curve is a consequence of the facts that at this curve,

$$\sigma(T,V_s) = S(T,V_s) - S_{id} = S - Nk\left\{\frac{5}{2} + \ln\left[v_s\left(\frac{mkT}{2\pi\hbar^2}\right)^{3/2}\right]\right\} = \text{const} \qquad (8)$$

and

$$c_v/\gamma = \text{const} \qquad (9)$$

where, in this case, $S$ is the lattice part of the entropy of the crystal, $c_v = C_v^l - 3R/2$, $C_v^l$ is the lattice part of the isochoric specific heat and $\gamma = (\partial P/\partial T)_V - R/V$. Actually, along the curve calculated, $c_v$ = const and $\gamma$ = const. Because of this, we can state with assurance that at the melting curve of $C_{60}$ fullerite (6), (7) evaluated using the Lindemann criterion (1), the entropy rule (8) is obeyed as well.

In conclusion, it may be said once more that the liquid phase of fullerites has not hitherto been observed experimentally and we have used the estimation for the melting point of C60 fullerite at normal pressure [36] $T_0 = 1500$ K. The experimental value of this temperature in the future may be somewhat different. In such a situation, it will be an easy matter to improve our calculations. It is inconceivable also that the evaluated curve (6), (7) lies on the metastable region of the solid and liquid phases. Nevertheless, the study of the laws of equilibrium between metastable phases of various materials is of significant interest to statistical thermodynamics [40–42] because sometimes two phases that are metastable with respect to a third phase may coexist in equilibrium with one another [40] and it is important to investigate their properties.

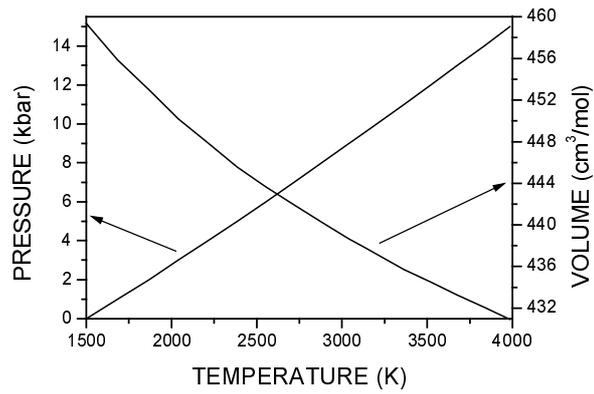

Fig. 1. The possible melting curve $P_m = P_m(T)$, $V_S = V_S(T)$ of $C_{60}$ fullerite.

**Acknowledgement** V. I. Zubov is grateful to Conselho Nacional de Desenvolvimento Científico e Tecnológico – CNPq (Brazil) for financial support.

## References

[1] H.W.Kroto, J.R.Heath, S.C.O'Brien, R.F.Curl and R.E.Smalley, Nature **318**, 162 (1985).

[2] W.Krätschmer, L.D.Lamb, K.Fostinoupolos and R.D.Huffman, Nature **347**, 354 (1990).

[3] H. W. Kroto, J. E. Fischer, and D. E. Cox (Eds.), The Fullerenes (Pergamon Press, Oxford, (1993).

[4] J. E. Fischer and P. A. Heiney, J. Phys. Chem. Solids 54, 1725 (1993).

[5] W.Krätschmer, Nuovo Cim. **107A**, 1077 (1994).

[6] A.V.Eletsky, B.M.Smirnov, Usp. Fiz. Nauk **165**, 977 (1995), in Russian.

[7] G.B.M.Vaughan, P.A.Heiney, J.E.Fischer, D.E.Luzzi, D.A. Rickets-Foot, A.R.McGhi, Y.-W.Hui, A.L.Smith, D.E.Cox, W.J.Romanow, B.H.Allen, N.Coustel, J.P.McCauley, Jr., and A.B.Smith III, Science **254**, 1350 (1991).

[8] M.A.Verheijen, H.Meekes, P.Bennema, J.L. de Boer, S. van Smaalen, G. van Tendoloo, S.Amelincks, S.Muto and J. van Landuyt, Chem. Phys. **166**, 287 (1992).

[9] W.I.F.David, R.M.Ibberson, D.T.S.Dennis, J.P.Hare and K.Prassidel, Europhys. Lett. **18**, 219 (1992).

[10] J. de Bruijn, A. Dworkin, H.Szwarz, J.Godard, R. Ceolin, C.Fabre and A. Rassat, Europhys. Lett. **24**, 551 (1993).

[11] H.A.Ludwig, W.H.Fietz, F.W.Hornung, K.Grube, B.Wagner and G.J.Burkhart, Z. Phys. **B96**, 179 (1994).

[12] V.Blank, M.Popov, S.Buga, V.Davydov, V.N.Denisov, A.N.Ivlev, B.N.Martin, V.Agafonov and R.Ceolin, Phys. Lett. **A188**, 281 (1994).

[13] R.Saito and G.Dresselhaus, Phys. Rev. **B49**, 2143 (1994).

[14] V.Yu.Markov, O.V.Boltanina and L.N.Sidorov, Zhurn. Fis. Khimii **75**, 5 (2001); Russ. J. Phys. Chem. **75**, 1 (2001).

[15] V.I.Zubov, J.F.Sanchez-Ortiz, J.N.Teixeira Rabelo and I.V.Zubov, Phys. Rev. **B55**, 6747 (1997).

[16] V.I.Zubov, N.P.Tretiakov, I.V.Zubov, J.B. Marques Barrio and J.N.Teixeira Rabelo, J. Phys. Chem. Solids **58**, 12, 2039 (1997).

[17] V.I.Zubov, N.P.Tretiakov and I.V.Zubov, *Eur. Phys. J.* **B17**, 629 (2000).

[18] N.W.Ashcroft, Europhys. Lett. **16**, 355 (1991); Nature **365**, 387 (1993).

[19] A.Cheng, M.L.Klein and C.Caccamo, Phys. Rev. Lett. **71**, 1200 (1993).

[20] M.H.J.Hagen, E.J.Meijer, G.C.A.M.Mooij, D.Frenkel and H.N.W.Lekkerkerker, Nature **365**, 425 (1993).

[21] L.Mederos and G.Navasqués, Phys. Rev. **B50**, 1301 (1994).

[22] C.Caccamo, Phys. Rev. **B51**, 3387 (1995).

[23] M.Nasegava and K.Ohno, Phys. Rev. **E54**, 3928 (1996).

[24] M.A.Abramo and G.Coppolino, Phys. Rev. **B58**, 2372 (1998).

[25] M.Nasegava, J. Chem. Phys. **111**, 5955 (1999).

[26] M.Nasegava and K.Ohno, J. Chem. Phys. **113**, 4315 (2000).

[27] A. L. C. Ferreira, J. M. Pacheco, and J. P. Prates-Ramalho, J. Chem. Phys. 113, 738 (2000).

[28] M.R.Statzer, P.A.Heiney, J.E.Fischer and A.R.McGhie, Phys. Rev. **B55**, 127 (1977).

[29] S.G.Kim and D.Tománek, Phys. Rev. Lett. **72**, 2418 (1994).

[30] M.A.Abramo, C.Caccamo, D.Costa and G.Pelicane, Europhys. Lett. **54**, 468 (2001).

[31] G.Leibfried, *Gittertheorie der mechanischen und thermischen Eigenschaften der Kristalle*, Springer-Verlag, Berlin, 1955, in German.

[32] M.Ross, Phys. Rev. **184**, 233 (1969).

[33] S.M.Stishov, Usp. Fiz. Nauk **114**, 3 (1974), in Russian.

[34] V.I.Zubov, Zh. Fiz. Khimii **55**, 2171 (1981), in Russian.

[35] Zheng-Hua Fang, J. Phys.: Cond. Mat. **8**, 7067 (1996).

[36] V.I.Zubov, N.P.Tretiakov, J.F.Sanchez and A.A.Caparica, Phys. Rev. **B53**, 12080 (1996).

[37 ]L.F.Girifalco, J. Phys. Chem. **96**, 858 (1992).

[38] V.I.Zubov and C.G.Rodrigues, Phys. stat. sol.(b) **222**, 471 (2000).

[39] V. B. Magalinskii and V. I. Zubov, phys. stat. sol. (b) 105, K139 (1981).

[40] V. P. Skripov, J. Non-Equilib. Thermodyn. 17, 193 (1992).

[41] S. C. Gulen, P. A. Thompson, and H. J. Cho, J. Fluid. Mech. 277, 163 (1994).

[42] V. P. Skripov and M. Z. Faizullin, High Temp. 37, 784 (1999).